\documentclass[conference]{IEEEtran}
\ifCLASSINFOpdf
   \usepackage[pdftex]{graphicx}
   \graphicspath{{../pdf/}{../jpeg/}}
   \DeclareGraphicsExtensions{.pdf,.jpeg,.png}
\else
   \usepackage[dvips]{graphicx}
   \graphicspath{{../eps/}}
   \DeclareGraphicsExtensions{.eps}
\fi

\usepackage{fixltx2e}
\usepackage{amsfonts}
\usepackage{amsmath,amssymb,amsfonts}
\usepackage{cite} 
\usepackage{graphicx}  
\usepackage{subfigure} 
\usepackage{amsmath,amssymb,amsfonts}
\usepackage{algorithmic}
\usepackage{textcomp}
\usepackage{algorithm}
\usepackage{multirow}
\usepackage{algorithmic}
\usepackage{bm}
\usepackage{cite}
\usepackage{float}
\usepackage{url}
\usepackage[pdftex]{graphicx}
\usepackage{amsmath} 
\usepackage{amssymb}
\usepackage{threeparttable}
\usepackage{tabularx}
\usepackage{url}

\usepackage{fancyhdr}

\begin{document}

\title{An AFDM-Based Integrated Sensing and Communications}


\author{\IEEEauthorblockN{Yuanhan Ni, Zulin Wang, Peng Yuan and Qin Huang}
	\IEEEauthorblockA{School of Electronic and Information Engineering, Beihang University, Beijing, 100191, China\\
		Email: \{yuanhanni, wzulin, yuanpeng9208\}@buaa.edu.cn, qhuang.smash@gmail.com (corresponding author)\\
}}


%


\maketitle


\begin{abstract}
This paper considers an affine frequency division multiplexing (AFDM)-based integrated sensing and communications (ISAC) system, where the AFDM waveform is used to simultaneously carry communications information and sense targets. To realize AFDM-based sensing functionality, two parameter estimation methods are designed to process echoes in the time domain and the discrete affine Fourier transform (DAFT) domain, respectively. It allows us to decouple delay and Doppler shift in the fast time axis and can maintain good sensing performance even in large Doppler shift scenarios. Numerical results verify the effectiveness of our proposed AFDM-based system in high mobility scenarios.

\end{abstract}

\begin{IEEEkeywords}
	Integrated sensing and communications, integrated waveform, affine frequency division multiplexing, discrete affine Fourier transform. 
\end{IEEEkeywords}

%
\IEEEpeerreviewmaketitle

\section{Introduction}

 



Next-generation wireless systems (beyond 5G/6G) are expected to maintain reliable communications in high mobility scenario, improve spectral and energy efficiencies significantly and support ubiquitous connections of everything \cite{IMT2021White}. The integrated sensing and communications (ISAC) technique is one of the key enablers for beyond 5G/6G due to its ability to improve spectral and energy efficiencies and obtain information about the environment \cite{liu2020joint,liu2022integrated}.

Integrated waveform design is one of the cornerstones of ISAC. One approach to design the integrated waveform is to endow new functionality to traditional waveforms that have been used in radar or communications systems \cite{ni2019high,Zhang2017A,sturm2011waveform,zeng2020joint}.
For example, the linear frequency modulation (LFM) waveform, a classical radar waveform, is used to embed communications information\cite{ni2019high,Zhang2017A}.  
Meanwhile, sensing functionality is added to the orthogonal frequency division multiplexing (OFDM) waveform that has been widely used in communications systems (LTE, 5G NR, WiFi, etc.) due to its robustness against multipath fading \cite{sturm2011waveform,zeng2020joint}. Specifically, in \cite{sturm2011waveform}, a symbol division-based method is proposed to suppress the sidelobes of radar image caused by random communications symbols and estimate parameters of targets, i.e., range and velocity. The symbol division operation is replaced with symbol conjugate multiplication by utilizing the cyclic cross-correlation (CCC) to avoid amplifying the noise background when symbols have non-constant modulus in the frequency domain \cite{zeng2020joint}. However, the peak-to-sidelobe level ratio (PSLR) of radar images may deteriorate severely in the high mobility scenarios due to the couple of delay and large Doppler shift in the fast time axis.


Apart from traditional waveforms, some new waveforms are investigated for both sensing and communications to improve the overall performance of ISAC systems \cite{Raviteja2018Interference,Ouyang2016Orthogonal}. The orthogonal time-frequency space (OTFS) waveform multiplexes information symbols in the delay-Doppler domain \cite{hadani2017orthogonal,Raviteja2018Interference}. From the perspective of communications, the OTFS has the ability to deal with large Doppler shift and obtain both time and frequency diversities in doubly selective channels. Moreover, it can achieve similar sensing performance with OFDM as shown in \cite{Gaudio2020On}. Meanwhile, a chirp-based multicarrier waveform is proposed, namely orthogonal
chirp-division multiplexing (OCDM) \cite{Ouyang2016Orthogonal,Omar2016Performance}.
An important advantage of chirp-based waveform is the potential to support full duplex operation \cite{wang2022towards}.
The OCDM multiplexes a set of orthogonal chirps that are complex exponentials with linearly varying instantaneous frequencies. Since each information is spanned the entire bandwidth, OCDM can achieve full diversity in frequency selective channels, which outperforms OFDM. The sidelobe level of  resulting radar image is slightly increased compared with the OFDM waveform \cite{Oliveira2020An,Oliveira2021MIMO}. However, OCDM can not achieve full diversity in doubly selective channels \cite{bemani2022affine}. 

Recently, another chirp-based waveform, namely affine frequency division multiplexing (AFDM), is proposed by multiplexing information symbols in the discrete affine Fourier transform (DAFT) domain \cite{bemani2021afdm, bemani2021affine,bemani2022affine}. The AFDM waveform can adapt to the channel profile by optimizing its parameters such that all paths can be separated in the DAFT domain. As a result, AFDM can achieve full diversity in doubly selective channels. The communications performances as well as processing algorithms have been investigated for AFDM in \cite{bemani2021afdm, bemani2021affine,bemani2022affine}. Results in \cite{bemani2022affine} show that compared with OTFS, AFDM has comparable communications performance in terms of bit error rate (BER) but with lower complexity and the advantage of less channel estimation overhead \cite{bemani2021affine,bemani2022affine}. Therefore, AFDM is considered as a potential candidate waveform for the ISAC system \cite{wang2022towards}.
However, it remains interesting to investigate the sensing performance of the AFDM waveform.

In this paper, we propose an AFDM-based ISAC system, where the AFDM waveform is used to simultaneously carry communications information of downlink user and sense information about the targets. Two methods are designed to estimate the range and velocity parameters of targets using random information symbols. The first method extracts parameters in the time domain utilizing the CCC with low complexity. The second method estimates parameters in the DAFT domain by compensating for the linear phase shift and estimating the cyclic shift of information symbols in the fast time axis. 
Benefiting from decoupling delay and Doppler shift in the fast time axis, the second method can enlarge the maximum unambiguous Doppler shift and maintain good sensing performance in large Doppler shift scenarios.
Numerical results show that our proposed AFDM-based system enjoys better PSLR performance than the OFDM-based system in high mobility scenarios.

%
%
\section{System Model}
We consider an AFDM-based ISAC system, where the ISAC base station (BS) transmits the AFDM waveform to the downlink user and simultaneously receives echoes reflected by the targets around it, as shown in Fig. 1.
\begin{figure}
	\centering	
	\includegraphics[width=1.7in]{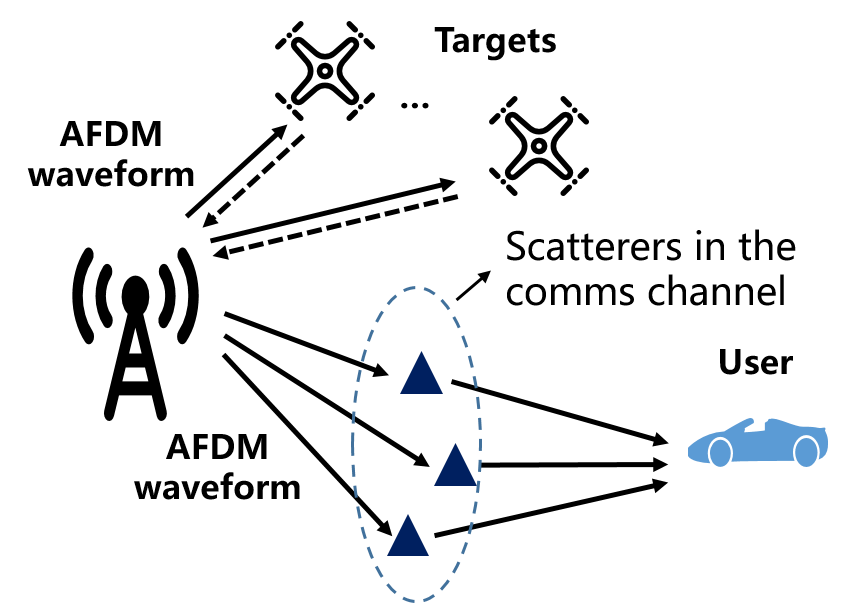}
	\caption{AFDM-based integrated sensing and communications system.
		\label{fg:AFDM_based_ISAC}}
\end{figure}

\subsection{Communications Signal Model}

In this subsection, we briefly review the basic concepts of AFDM proposed in \cite{bemani2022affine, bemani2021afdm, bemani2021affine}. Let $\bm{x}$ denote an $N {\times} 1$ vector of quadrature amplitude modulation (QAM) symbols. The $N$ points inverse DAFT (IDAFT) is performed to map $\bm{x}$ to the time domain as
\begin{equation}
\bm{s}\left[ {n} \right] = \frac{1}{{\sqrt N }}\sum\nolimits_{m = 0}^{N - 1} {\bm{x}\left[ {m} \right]} {e^{j2\pi \left( {{c_1}{n^2} + \frac{1}{N}mn + {c_2}{m^2}} \right)}} ,
\end{equation}
where $c_1$ and $c_2$ are the AFDM parameters, and $n = 0, \ldots ,N {-} 1$. Then, a chirp-periodic prefix (CPP) is added with a length of $N_{cp}$. 

After transmission over the channel, discarding CPP and performing $N$ points DAFT, the received sample matrix in the DAFT domain can be written in the matrix form as
\begin{equation}
\bm{y} = {\bm{H}_{\rm eff}}\bm{x} + \bm{{\tilde w}},
\end{equation}
where $\bm{{\tilde w}}\sim \mathcal {CN}\left( {0\text{,} \sigma_{c} ^2\bm{I}} \right)$ is an additive Gaussian noise vector and ${\bm{H}_{\rm eff}} {=} {\bm{\Lambda} _{{c_2}}}\bm{F}{\bm{\Lambda} _{{c_1}}}{\bm{H}_c}\bm{\Lambda} _{{c_1}}^{\rm{H}}{\bm{F}^{\rm{H}}}\bm{\Lambda} _{{c_2}}^{\rm{H}}$ with $\bm{F}$ being the discrete Fourier transform (DFT) matrix\cite{bemani2022affine}, ${\bm{\Lambda} _{{c_i}}} {=}  diag\left( {{e^{ - j2\pi c_i{n^2}}},n = 0, \ldots ,N {-} 1} \right)$, ${\bm{H}_c}$ being the matrix representation of the communications channel in the time domain, and $(\cdot)^{\rm H}$ denoting Hermitian transpose.
	
\subsection{Radar Signal Model}

This paper considers that the ISAC BS uses an AFDM frame with $N_{sym}$ AFDM symbols to sense the targets around it. We derive the AFDM-based signal model of target echoes. 

Let $\bm{X} {=} \left[ {{\bm{x}_0}, \ldots ,{\bm{x}_{N_{sym} - 1}}} \right]$ and $\bm{S} {=} \left[ {{\bm{s}_0}, \ldots ,{\bm{s}_{N_{sym} - 1}}} \right]$ denote the $N {\times} N_{sym}$ blocks of QAM symbols and modulated signals in time domain, respectively. After adding CPP and parallel to serial conversion, the signal to be transmitted is given by $\bm{\tilde s} {\in} {\mathbb{C}^{\left( {N + {N_{cp}}} \right){N_{sym}} \times 1}}$. Consider $P$ targets. The scattering
coefficient, range, velocity, time delay and Doppler frequency of the $i$-th target are denoted by ${h_i}$, ${R_i}$, ${v_{rel,i}}$, ${\tau _i} {=} {{2{R_i}} \mathord{\left/
		{\vphantom {{2{R_i}} c}} \right.
		\kern-\nulldelimiterspace} c}$, ${f_{d,i}} {=} {{2{v_{rel,i}}{f_c}} \mathord{\left/
		{\vphantom {{2{v_{rel,i}}{f_c}} c}}\right.
		\kern-\nulldelimiterspace} c}$ with $c$ and ${f_c}$ being the speed of light and the frequency of carrier, respectively, and $i = 1, \ldots ,P$. The received target echo in the time domain is \cite[Eq. (6)]{wu2022integrating}
\begin{equation}
\bm{r}\left[ n \right] = \sum\nolimits_{i = 1}^P {{{\tilde h}_i}} \bm{s}\left[ {n - {l_i}} \right]{e^{j2\pi {f_i}n}} + {\bm{w}_r}\left[ n \right],
\end{equation}
where $n {=} 0, \ldots ,{\left( {N {+} {N_{cp}}} \right){N_{sym}}} {-} 1$, ${{\bm{w}_r}\left[ n \right] \sim \mathcal {CN}\left( {0\text{,} \sigma_{r} ^2} \right)}$ is the noise, ${{\tilde h}_i} {=} {h_i}{e^{ - j2\pi {f_{d,i}}{\tau _i}}}$, ${l_i} {=} {{{\tau _i}} \mathord{\left/
		{\vphantom {{{\tau _i}} {{t_s}}}} \right.
		\kern-\nulldelimiterspace} {{t_s}}}$, ${f_i} {=} {f_{d,i}}{t_s}$, and ${t_s}$ denotes the sampling interval. Following \cite{sturm2011waveform,zeng2020joint,bemani2022affine}, it is assumed that $N_{cp}$ is greater than the maximum delay $l_{max}$. We define ${\nu _i} = N{f_i} = \frac{{{f_{d,i}}}}{{\Delta f}} = {\alpha _i} + {a_i}$, where ${\nu _i} \in \left[ { - {\nu _{\max }}, {\nu _{\max }}} \right]$ is the Doppler shift normalized with respect to the 
	subcarrier spacing ${\Delta f}$, ${\alpha _i} {\in} \left[ { - {\alpha _{\max }},{\alpha _{\max }}} \right]$ is its integer part, and ${a_i} {\in} \left( { - \frac{1}{2},\frac{1}{2}} \right]$ is its fractional part\cite{bemani2022affine}. 
	
After serial to parallel conversion and discarding CPP, the received target echo can be expressed as
\begin{equation}
\bm{R}\left[ {n,k} \right] {=} \sum\limits_{i = 1}^P {{{\tilde h}_i}} \bm{S}\left[ {n {-} {l_i},k} \right]{e^{j2\pi {f_i}\left[ {n{\rm{ + }}\left( {N + {N_{cp}}} \right)k} \right]}} {+} {\bm{W}_r}\left[ {n,k} \right],
\end{equation}
where ${\bm{W}_r}$ denotes noise matrix, $n = 0, \ldots ,N {-} 1$, and $k = 0, \ldots ,N_{sym} {-} 1$.
Performing $N$ points DAFT on each column of $\bm{R}$, the DAFT domain samples matrix $\bm{Y}_r$ can be given by
\begin{align}
&\bm{Y}_r\left[ {m,k} \right] {=} \frac{1}{N}\sum\limits_{n = 0}^{N - 1} {\bm{R}\left[ {n,k} \right]{e^{ - j2\pi \left( {{c_1}{n^2} {+} \frac{1}{N}mn {+} {c_2}{m^2}} \right)}}},
\end{align}
where $m = 0, \ldots ,N {-} 1$ and $k = 0, \ldots ,N_{sym} {-} 1$.

As a result, motivated by \cite[Eq. (26)]{bemani2022affine}, the input-output relation in DAFT domain can be written in the matrix form as
\begin{equation} \label{eq:in_out_relation_1}
\bm{Y}_r =\sum\nolimits_{i = 1}^P {{{\tilde h}_i}{\bm{H}_i}\bm{X}\bm{D}_i} + \bm{\tilde W}_r,
\end{equation}
where $\bm{D}_i= diag\left( {{e^{ - j2\pi {f_i}\left( {N + {N_{cp}}} \right)k}}} \right)$, $k {=} 0, \ldots ,{N_{sym}} {-} 1$, ${\bm{H}_i}\left[ {p,q} \right] = \frac{1}{N}{e^{j\frac{{2\pi }}{N}\left( {N{c_1}l_i^2 - ql_i + N{c_2}\left( {{q^2} - {p^2}} \right)} \right)}}\mathcal{F}_i\left( {p,q} \right)$ and $\bm{\tilde W}_r={\bm{\Lambda} _{{c_2}}}\bm{F}{\bm{\Lambda} _{{c_1}}}\bm{W}_r$. Different from \cite{bemani2022affine}, here ${\mathcal{F}_i}\left( {p,q} \right) {=} \frac{{{e^{ - j2\pi \left( {p - q - {v_i} + 2N{c_1}{l_i}} \right)}} - 1}}{{{e^{ - j\frac{{2\pi }}{N}\left( {p - q - {v_i} + 2N{c_1}{l_i}} \right)}} - 1}}$. 
Similar to \cite{bemani2022affine}, in the integer normalized Doppler shift case, i.e., $a_i{=}0$, there is only one non-zero element in each row of $\bm{H}_i$, i.e., the element of ${\bm{H}_i}$ at row $p$ and column $q$ can be written as 
\begin{equation}
{\bm{H}_i}\left[ {p,q} \right] = \left\{ {\begin{array}{*{20}{c}}
	\hspace{-0.2cm}{{e^{j\frac{{2\pi }}{N}\left( {N{c_1}l_i^2 - ql_i + N{c_2}\left( {{q^2} - {p^2}} \right)} \right)}}},&{q = {{\left( {p + lo{c_i}} \right)}_N}},\\
	0,&{otherwise},
	\end{array}} \right. 
\end{equation}
where $lo{c_i} {=} {\left( {2N{c_1}{l_i} {-} {\alpha _i}} \right)_N}$ and ${\left( \cdot \right)_N}$ is the modulo $N$ operation. Hence, the input-output relation can be rewritten as
\begin{align}\label{eq:in_out_relation_2}
&\bm{Y}_r\left[ {p,k} \right]  = \sum\limits_{i = 1}^P {{\tilde h}_i}{e^{  j2\pi {f_i}\left( {N + {N_{cp}}} \right)k}} \nonumber\\ 
	&\times {e^{j\frac{{2\pi }}{N}\left( {N{c_1}l_i^2 - q_il_i + N{c_2}\left( {{{q}_i^2} - {p^2}} \right)} \right)}}\bm{X}\left[ {q_i,k} \right]{+} \bm{\tilde W}_r\left[ {p,k} \right],
\end{align}
where $p {=} 0, \ldots ,N {-} 1$, ${q_i {=} {{\left( {p + lo{c_i}} \right)}_N}}$ and $k {=} 0, \ldots ,$ $N_{sym} {-} 1$.
In the fractional normalized Doppler shift case, 
there are $2k_v {+} 1$ non-zero elements and the peak is still at ${q {=} {{\left( {p + lo{c_i}} \right)}_N}}$ in each row of $\bm{H}_i$. 

\subsection{Communications Processing}
The communications processing algorithms for AFDM, i.e., channel estimation, equalization and symbol detection, have been studied in \cite{yin2022pilot,bemani2021afdm, bemani2021affine,bemani2022affine}. In the AFDM-based ISAC system, the downlink user can exploit these methods to process received AFDM waveform and obtain the communications information.
In this paper, we mainly focus on estimating the parameters of targets based on the AFDM waveform.

\section{AFDM-Based Parameter Estimation Methods} \label{Sec_3}
In this section, we propose two parameter estimation methods based on the AFDM waveform to estimate the range and velocity of targets in the time domain and the DAFT domain, respectively.

\subsection{Parameter Estimation in Time Domain}
We exploit the Fast Cyclic Correlation Radar (FCCR) algorithm \cite{zeng2020joint} to estimate the range and velocity of targets. Specifically, each column of the received signal matrix $\bm{R}$ and the transmit matrix $\bm{S}$ in the time domain undergoes the $N$ points DFT, which results in two $N {\times} N_{sym}$ matrices in the frequency domain. 
Then, the element-based conjugate multiplication is performed on the resulting matrices. The $N$ points inverse DFT (IDFT) is computed for each column of matrix and the $N_{sym}$ points DFT is calculated for each row of matrix. After these
steps, a two-dimensional (2-D) range-Doppler matrix (RDM) is produced, which represents a 2-D radar image in range and Doppler.

As mentioned in \cite{sturm2011waveform}, the processing gain $G_p{=}NN_{sym}$, and the maximum unambiguous Doppler shift is given by $\pm{1 \mathord{\left/
{\vphantom {1 {\left( {2{T_{AFDM}}} \right)}}} \right.
\kern-\nulldelimiterspace} {\left( {2{T_{AFDM}}} \right)}}$, where $T_{AFDM}$ denotes the duration of a AFDM symbol including CPP. 

\subsection{Parameter Estimation in DAFT Domain}
We start from analyzing the input-output relation in the DAFT domain and then propose a method to estimate the range and velocity of targets in the DAFT domain. This method is derived utilizing the input-output relation in the integer normalized Doppler shift case, and numerical results show that it is also suitable for the fractional normalized Doppler shift case. 

Following {\cite{bemani2022affine}, $c_1$ is set as ${c_1} = \frac{{2\left( {{\alpha _{\max }} + {k_v}} \right)+1}}{{2N}}$ and $c_2$ is set to be a rational number sufficiently smaller than $\frac{1}{{2N}}$. Thus, if $c_2$ is small enough, the value of $N{c_2}\left( {{{q_i}^2} - {p^2}} \right)$ will approach zero. Let ${q_i {=} {{\left( {p + lo{c_i}} \right)}_N}} {=} p {+} d_i$, i.e., $d_i {=} {\left( {p + lo{c_i}} \right)}_N {-} p$. Equation (\ref{eq:in_out_relation_2}) can be rewritten as 
\begin{align}\label{eq:in_out_relation_3}
\bm{Y}_r\left[ {p,k} \right] \approx & \sum\limits_{i = 1}^P {{\zeta _i}{e^{ j2\pi {f_i}\left( {N {+} {N_{cp}}} \right)k}}{e^{ - j\frac{{2\pi }}{N}{l_i}p}}\bm{X}\left[ {p {+} {d_i},k} \right]} \nonumber\\
&{+} \bm{\tilde W}_r\left[ {p,k} \right],
\end{align}
where ${\zeta _i} = {{\tilde h}_i}{e^{j\frac{{2\pi }}{N}\left( {N{c_1}l_i^2 - {d_i}{l_i}} \right)}}$. We can see from (\ref{eq:in_out_relation_3}) that in each row of $\bm{Y}_r$, i.e. fixing $p$, there is the linear phase shift between the information symbols $\bm{X}\left[ {p,:} \right]$ along the $k$-axis, which is caused by the Doppler shift.
In each column of $\bm{Y}_r$, i.e. fixing $k$, there are the linear phase shift between the information symbols $\bm{X}\left[ {:,k} \right]$ and the cyclic shift of information symbols $\bm{X}\left[ {:,k} \right]$. The former is caused by the delay $l_i$, and the latter is caused by the delay $l_i$ and the integer part of normalized Doppler shift, i.e., ${\alpha _i}$. Due to this couple of linear phase shift and cyclic shift along the $p$-axis, the sidelobes of the radar image obtained by the FCCR algorithm will be severely deteriorated when the Doppler shift is large. Numerical results will verify this conclusion. 

To this end, we propose a parameter estimation method in the DAFT domain, which consists of four steps, to enjoy good performance in large Doppler scenarios. Specifically, in the first step, we compensate for the linear phase shift along the $p$-axis caused by the delay $l_i$. The index of the cyclic shift of information symbols $\bm{X}$ and the linear phase shift along the $k$-axis caused by the Doppler shift are estimated by the matched filer in the DAFT domain and the DFT operation, respectively, in the second step. In the third step, the delay and Doppler shift are extracted from the resulting matrices using our derived relationships of delay and Doppler shift to the peak index. In the fourth step, the extracted integer and fractional normalized Doppler shifts are combined to enlarge the maximum unambiguous Doppler shift.


\subsubsection{Compensating for linear phase shift} We first remove the effect of the linear phase shift in each column of $\bm{Y}_r$ by actively compensating for it. According to the above assumption that the CPP length $N_{cp}$ is greater than the maximum delay $l_{max}$, the delay $l$ is satisfied $0 \le {l} < {N_{cp}}$. For each delay $l$, we generate a compensation matrix $\bm{L}_l = diag\left( {{e^{j\frac{{2\pi }}{N}{l}p}},p {=} 0, \ldots ,N - 1} \right)$ and multiply it by $\bm{Y}_r$, i.e., $\bm{Z}_l = \bm{L}_l\bm{Y}$ and 
\begin{align}
\bm{Z}_l\left[ {p,k} \right] = &\sum\limits_{i = 1}^P {{\zeta _i}{e^{ j2\pi {f_i}\left( {N + {N_{cp}}} \right)k}}{e^{j\frac{{2\pi }}{N}\left( {l - {l_i}} \right)p}}\bm{X}\left[ {p + {d_i},k} \right]}\nonumber\\
&{+} \bm{\tilde W}_r\left[ {p,k} \right], \ l {=} 0, \ldots ,N_{cp} - 1.
\end{align} 
This process results in $N_{cp}$ matrices.

\subsubsection{Matched filer in DAFT domain and DFT} Next, we exploit the matched filter in the DAFT domain to estimate the index of the cyclic shift of information symbols $\bm{X}$. Specifically, for $l {=} 0, \ldots ,N_{cp} - 1$, each column of $\bm{Z}_l$ and $\bm{X}$ undergoes a $N$ points DFT. Then, the element-based conjugate multiplication is performed. Finally, the $N$ points IDFT is computed for each column and the $N_{sym}$ points DFT is calculated for each row\footnote{The result of DFT is circularly shifted by ${{{{N_{sym}}} \mathord{\left/
				{\vphantom {{{N_{sym}}} 2}} \right.
				\kern-\nulldelimiterspace} 2}}$ positions.}. The resulting matrix can be written in the matrix form as
\begin{equation}
\bm{W}_l = {\bm{F}^{\rm{H}}}\left( {{{\left( {\bm{F}\bm{L}_l\bm{Y}_r} \right)}^*} \odot \left( {\bm{F}\bm{X}} \right)} \right)\bm{F},
\end{equation}
where $(\cdot)^{*}$ and $\odot$ are conjugate and Hadamard product, respectively. 

\subsubsection{Extracting delay and Doppler shift} After the previous steps, we get $N_{cp}$ matrices that contain the delay and Doppler information of $P$ targets. Next, we investigate how to extract these information from these $N_{cp}$ matrices.
When $l {=} l_i$, ${e^{j\frac{{2\pi }}{N}\left( {l - {l_i}} \right)p}} {=} 1$ for $p {=} 0, \ldots ,N - 1$, i.e., the effect of linear phase shift caused by the delay is removed, and the resulting matrix $\bm{W}_l$ will contain one or more peaks (depending on how many targets have this delay and different Doppler shift). 
We can use the existing constant false alarm rate (CFAR) algorithm to detect the existence of targets \cite{kronauge2013fast}. 
If the magnitude of peak of $\bm{W}_l$, $l {=} 0, \ldots ,N_{cp} - 1$, exceeds the threshold, there is a target whose corresponding delay estimation is $\hat l_i = l$. 
Let ${\bar p_i}$ and ${\bar k_i}$ denote the indexes of row and column of the peak exceeding the threshold in $\bm{W}_l$. 
As mentioned earlier, the index of the cyclic shift of $\bm{X}$ is caused by $l_i$ and $\alpha _i$. Therefore, we can get the estimation of integer part of normalized Doppler shift, given by
\begin{equation}
{\hat \alpha _i} = 2N{c_1}{\hat l_i} {-} {\left( {1 - {{\bar p}_i}} \right)_N}.
\end{equation}
Similar to the time domain method in Sec. \ref{Sec_3}. A, the maximum unambiguous Doppler shift that ${\bar k_i}$ can represent is $\pm{1 \mathord{\left/
		{\vphantom {1 {\left( {2{T_{AFDM}}} \right)}}} \right.
		\kern-\nulldelimiterspace} {\left( {2{T_{AFDM}}} \right)}} {=}  \pm \frac{1}{2}\Delta f'$, where $\Delta f' {=} {1 \mathord{\left/
		{\vphantom {1 {{T_{AFDM}}}}} \right.
		\kern-\nulldelimiterspace} {{T_{AFDM}}}}$. 
We define $\nu'_i {=} \frac{{{f_{d,i}}}}{{\Delta f'}} {=} {\beta _i} {+} {b_i}$, where ${\nu' _i} {\in} \left[ { - {\nu' _{\max }}, {\nu' _{\max }}} \right]$ is the Doppler shift normalized with respect to the maximum unambiguous Doppler shift ${\Delta f'}$, ${\beta _i} {\in} \left[ { - {\beta _{\max }},{\beta _{\max }}} \right]$ is its integer part, and ${b_i} {\in} \left( { - \frac{1}{2},\frac{1}{2}} \right]$ is the fractional part. Consequently, the estimation of the fractional Doppler shift ${b_i}$ is given by
\begin{equation}
{\hat b_i} = \frac{{\left( {{{\bar k}_i} - 1 - {{{N_{sym}}} \mathord{\left/
					{\vphantom {{{N_{sym}}} 2}} \right.
					\kern-\nulldelimiterspace} 2}} \right)}}{{{N_{sym}}}}.
\end{equation} 

\subsubsection{Combining integer and fractional Doppler shift}
Now, we have extracted the delay $l_i$, the integer Doppler shift $\alpha _i$ (normalized with ${\Delta f}$) and the fractional Doppler shift ${b_i}$ (normalized with ${\Delta f'}$) from matrices $\bm{W}_l$, $l {=} 0, \ldots ,N_{cp} - 1$. If integer and fractional Doppler shifts can be combined, the maximum unambiguous Doppler shift will be significantly increased for a given $T_{AFDM}$. Note that the values of $\alpha _i$ and ${b_i}$ cannot be directly added because they are the integer and fractional parts of two different normalized Doppler shifts $\nu$ and $\nu'$, respectively. According to the definitions of $\alpha _i$ and ${b_i}$, we can get 
\begin{equation}
{{\hat f}_{d,i}} = \left( {\hat \alpha _i  + \hat a _i} \right)\Delta f = \left( {\hat \beta _i + \hat b _i} \right)\Delta f',
\end{equation}
and
\begin{align}\label{eq:a_1}
\hat a _i &= \frac{{\Delta f'}}{{\Delta f}}\left( {\hat \beta _i + \hat b _i} \right) - \hat \alpha_i  \nonumber\\
& = \frac{N}{{N + {N_{cp}}}}\hat \beta _i + \frac{N}{{N + {N_{cp}}}}\hat b _i- \hat \alpha _i.
\end{align}
We can see that $\hat a _i$ is a function of $\hat \beta _i$.
According to the condition ${a_i} \in \left( { - \frac{1}{2},\frac{1}{2}} \right]$, we can get $\hat \beta_i  \in \left( {\frac{{N + {N_{cp}}}}{N}\hat \alpha_i  - \hat b_i - 0.5\frac{{N + {N_{cp}}}}{N},\frac{{N + {N_{cp}}}}{N}\hat \alpha_i  - \hat b_i + 0.5\frac{{N + {N_{cp}}}}{N}} \right]$. The minimum and maximum values of $\hat \beta_i$ are ${{{\hat \beta }_{i,\min }} = \left\lceil {\frac{{N + {N_{cp}}}}{N}\hat \alpha_i  - \hat b_i - 0.5\frac{{N + {N_{cp}}}}{N}} \right\rceil }$ and ${{\hat \beta }_{i,\max }} = \left\lfloor {\frac{{N + {N_{cp}}}}{N}\hat \alpha_i  - \hat b_i + 0.5\frac{{N + {N_{cp}}}}{N}} \right\rfloor $, respectively, where $\left\lceil \cdot \right\rceil $ and $\left\lfloor \cdot \right\rfloor$ denote the cell and floor functions, respectively. When $0 \le {N_{cp}} < N$, there are only two cases, either ${{\hat \beta }_{i,\max }} {=} {{\hat \beta }_{i,\min }}$ or ${{\hat \beta }_{i,\max }} {=} {{\hat \beta }_{i,\min }} {+} 1$. If ${{\hat \beta }_{i,\max }} {=} {{\hat \beta }_{i,\min }}$, let ${{\hat \beta_i }} {=} {{\hat \beta }_{i,\max }}$ and substitute it into (\ref{eq:a_1}) to compute $\hat a _i$. If ${{\hat \beta }_{i,\max }} {=} {{\hat \beta }_{i,\min }} {+} 1$, we introduce a criterion to judge whether ${{\hat \beta_i }}$ should be equal to ${{\hat \beta }_{i,\max }}$ or ${{\hat \beta }_{i,\min }}$, i.e.,
\begin{equation}
\hat \beta_i  = \left\{ {\begin{array}{*{20}{c}}
	{{{\hat \beta }_{i,\min }},}&{{\bm{W}_{\hat l_i}}\left[ {{{\bar p}_i} - 1,{{\bar k}_i}} \right] > {\bm{W}_{\hat l_i}}\left[ {{{\bar p}_i} + 1,{{\bar k}_i}} \right],}\\
	{{{\hat \beta }_{i,\max }},}&{otherwise,}
	\end{array}} \right.
\end{equation}
where ${\bm{W}_{\hat l_i}}\left[ {{{\bar p}_i}{-}1,{{\bar k}_i}} \right]$ and ${\bm{W}_{\hat l_i}}\left[ {{{\bar p}_i}{+}1,{{\bar k}_i}} \right]$ are two elements next to the peak (i.e., ${\bm{W}_{\hat l_i}}\left[ {{{\bar p}_i},{{\bar k}_i}} \right]$) along the $p$-axis, as shown in Fig. \ref{fg:ahead_after_peak}. 
\begin{figure}
	\centering
	\subfigure[]{
		\includegraphics[width=1.5in]{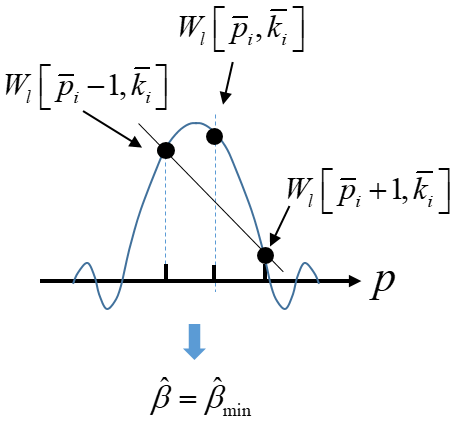}
	}
	\subfigure[]{
		\includegraphics[width=1.5in]{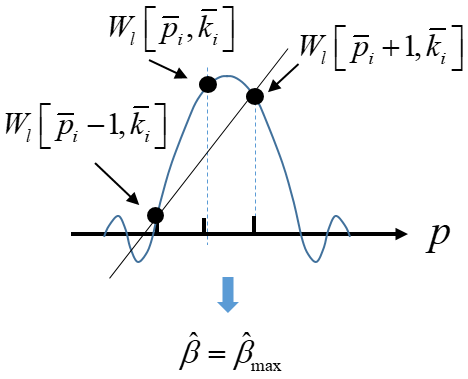}
	}	
	\vspace*{-5pt} 
	\caption{Judgment criterion: (a) if ${\bm{W}_{\hat l_i}}\left[ {{{\bar p}_i}{-}1,{{\bar k}_i}} \right] > {\bm{W}_{\hat l_i}}\left[ {{{\bar p}_i}{+}1,{{\bar k}_i}} \right]$, ${{\hat \beta_i }} {=} {{\hat \beta }_{i,\min }}$; (b) if ${\bm{W}_{\hat l_i}}\left[ {{{\bar p}_i}{-}1,{{\bar k}_i}} \right] \le {\bm{W}_{\hat l_i}}\left[ {{{\bar p}_i}{+}1,{{\bar k}_i}} \right]$, ${{\hat \beta_i }} {=} {{\hat \beta }_{i,\max }}$.  
		\label{fg:ahead_after_peak}}
	\vspace*{-5pt} 
\end{figure}


\section{Numerical Simulations and Analyses}
In this section, numerical results based on Monte Carlo simulations are presented. Following \cite{sturm2011waveform}, our simulation parameters are listed in Table \ref{tab:parameter_simu}. The information bits are modulated by 16-QAM. Let ${\rm SNR}_{\rm sig}$ denote the signal-to-noise ratio (SNR) of received signal, and we set $\alpha _{max}$ = 2 and ${k_v}$ = 4 \cite{bemani2022affine}. As a comparison, the performance of OFDM waveform processed by the symbol division method is also presented \cite{sturm2011waveform}. 

\renewcommand\arraystretch{1.2}
\begin{table}
	\centering
	\caption{Parameters in our simulations}\label{tab:parameter_simu}
	\begin{threeparttable}
		\begin{tabular}{|c|c|c|}
			\hline
			\textbf{Symbol}                       &\textbf{Parameter}                       & \textbf{Value}     \\ \hline
			$f_c$ &  Carrier frequency                      & $24$ GHz     \\ \hline
			$N$ &  Number of subcarriers                      & $4096$     \\ \hline
			$N_{sym}$ &  Number of AFDM symbols                      & $256$     \\ \hline
			$B$ &  Signal Bandwidth                       & $93.1$ MHz  \\ \hline		
			$\Delta _f$ &  Subcarrier spacing                      & $22.729$ kHz  \\ \hline	
			$T$ &  AFDM symbol duration                      & $44$ $\rm \mu$s  \\ \hline	
			$N_{cp}$ &  Chirp-periodic prefix length                      & $256$ ${\rm \mu}$s  \\ \hline
			$T_{cp}$ &  Chirp-periodic prefix duration                      & $2.75$ ${\rm \mu}$s  \\ \hline
			$T_{AFDM}$ &  Total AFDM symbol duration                      & $46.75$ ${\rm \mu}$s  \\ \hline
			$\Delta _R$ &  Range resolution                      & $1.61$ m  \\ \hline
			$\Delta _v$ &  Velocity resolution                      & $0.52$ m/s  \\ \hline
			$G _{p}$ &  Processing gain                & $60.2$ dB  \\ \hline
		\end{tabular}
	\end{threeparttable}
\end{table}

Figure \ref{fg:radar_range_profile} shows the normalized radar images
obtained by the OFDM and AFDM waveforms with different Doppler shift. ``OFDM-symbol-division'', ``Pro-time-domain'' and ``Pro-DAFT-domain'' represent the OFDM-based symbol division method, our proposed AFDM-based methods in the time domain and the DAFT domain, respectively.
For the symbol division method and our time-domain method, this radar image denotes radar range profile whose peak indicates the range of target. For our DAFT-domain method, the peak of this radar image simultaneously contains the information of the range and velocity of target. Therefore, the location of the peak obtained by our DAFT-domain method differs from that of the other two methods, as shown in Fig. \ref{fg:radar_range_profile}. When $f_d{=}0.1\Delta_f$, our time-domain and DAFT-domain methods offer similar PSLRs, which are slightly lower than that of the symbol division method. When $f_d{=}0.98\Delta_f$, the PSLRs of our time-domain method and the symbol division method deteriorate severely. However, our DAFT-domain method still maintains good PSLR performance.

\begin{figure}
	\centering
	\subfigure[$f_d{=}0.1\Delta_f$]{
		\includegraphics[width=2.0in]{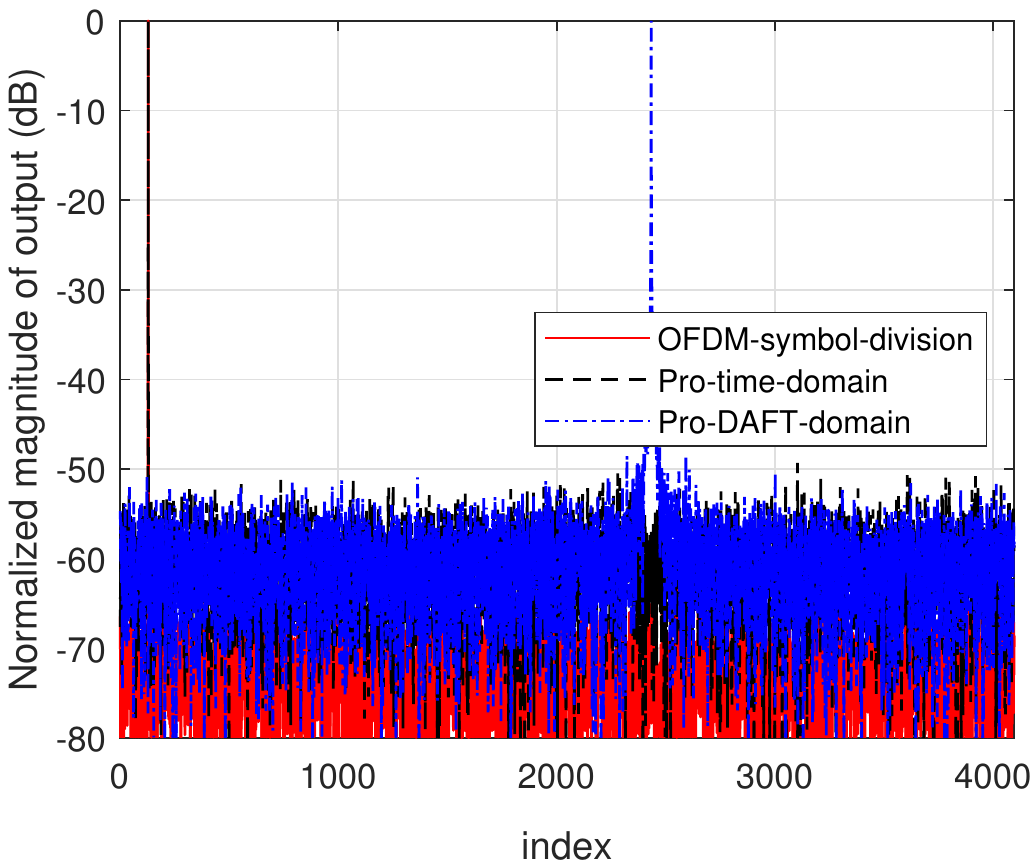}
	}
	\subfigure[$f_d{=}0.98\Delta_f$]{
		\includegraphics[width=2.0in]{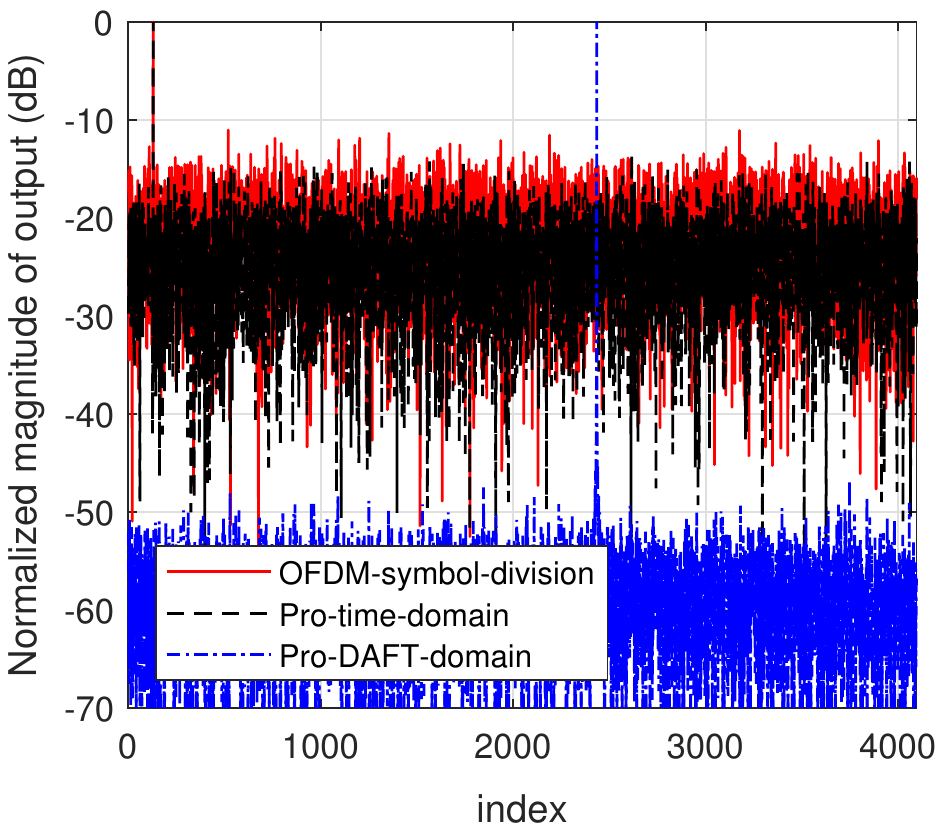}
	}	
	\vspace*{-5pt} 
	\caption{Calculated radar image for OFDM and AFDM waveforms with ${\rm SNR}_{\rm sig} {=} 10$ dB, $P{=}1$ and $l{=}128$.  
		\label{fg:radar_range_profile}}
	\vspace*{-5pt} 
\end{figure}

\begin{figure}
	\centering
	\includegraphics[width=2.0in]{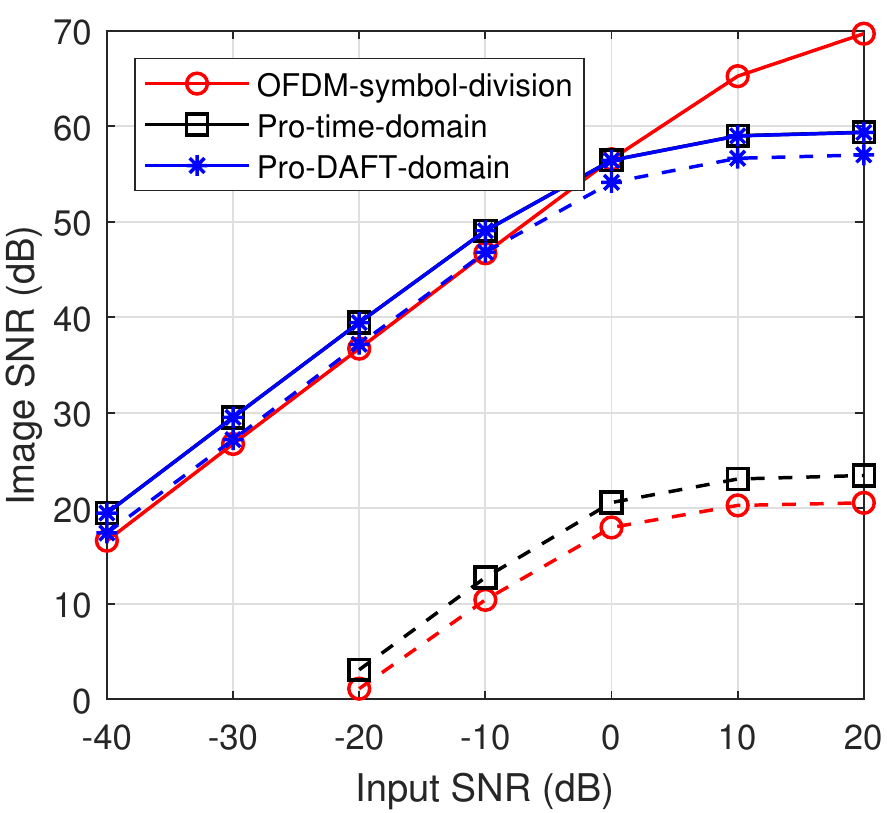}
	\caption{Image SNR versus the SNR of received signal with $P{=}1$ and $l{=}128$.
		\label{fg:Image_SNR_vs_inputSNR}}
\end{figure}

Figure \ref{fg:Image_SNR_vs_inputSNR} illustrates the corresponding image SNR, which is the ratio between the peak caused by target and the average noise level in the 2-D radar image \cite{sturm2011waveform}, versus the SNR of received signal. The solid and dashed lines represent the cases of $f_d{=}0.1\Delta_f$ and $f_d{=}0.98\Delta_f$, respectively.
It can be observed that in the case of $f_d{=}0.1\Delta_f$, the available image SNRs decrease almost linearly with ${\rm SNR}_{\rm sig}$ for all methods for a ${\rm SNR}_{\rm sig}$ below 0 dB.
There appears a saturation of image SNR starting approximately at the ${\rm SNR}_{\rm sig}$ of 10 dB for our proposed methods. The symbol division method can output higher image SNR at high ${\rm SNR}_{\rm sig}$ regions. 
In the case of $f_d{=}0.98\Delta_f$, there appears a saturation of image SNR for all methods, and the image SNRs severely decrease for our time-domain method and the symbol division method. However, the image SNR of our DAFT-domain method decreases slightly.
Furthermore, our two methods outperform the symbol division method. 

\begin{figure}
	\centering
	\includegraphics[width=2.0in]{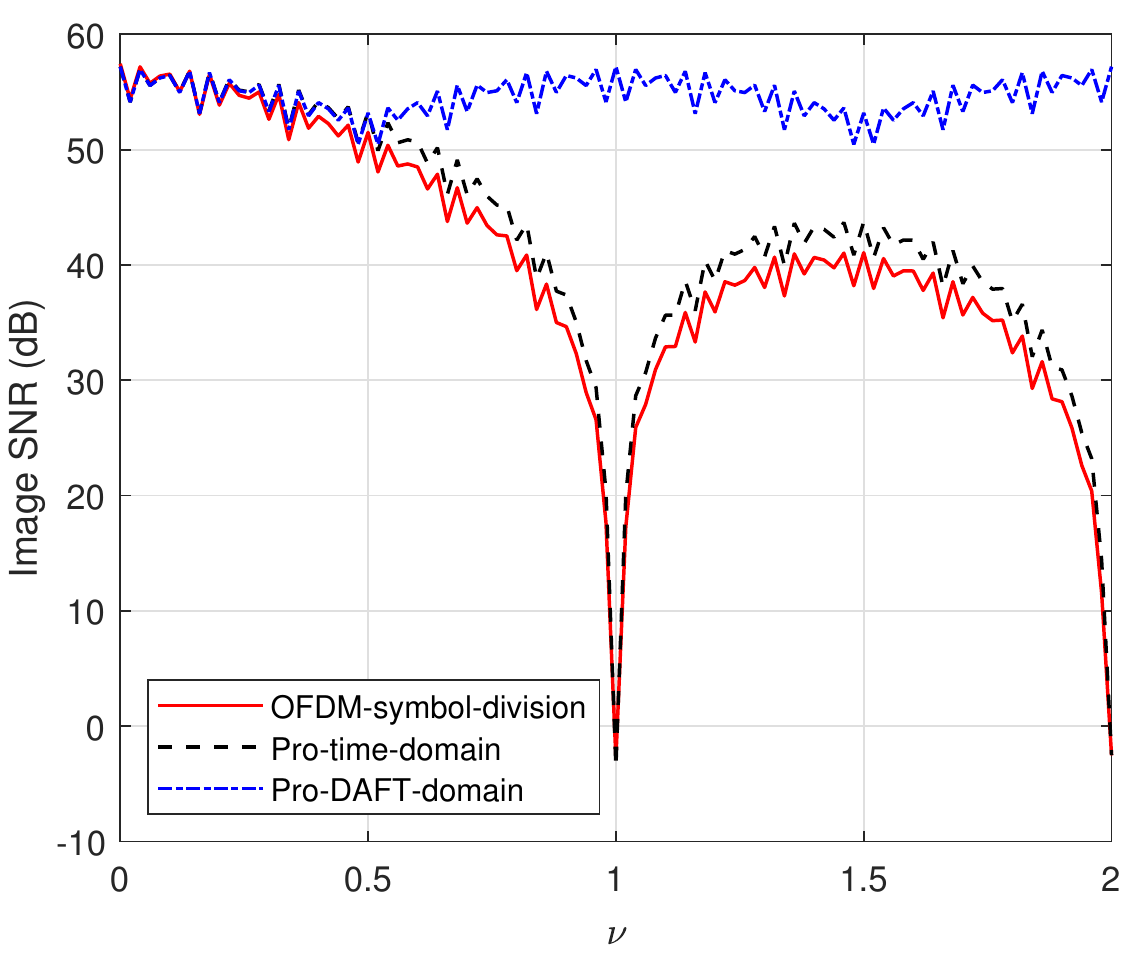}
	\caption{Image SNR versus the normalized Doppler shift $\nu$ with ${\rm SNR}_{\rm sig} {=} 0$ dB, $P{=}1$ and $l{=}128$.
		\label{fg:Image_SNR_vs_Doppler}}
\end{figure}
We compare the image SNR versus normalized Doppler shift $\nu$, as shown in Fig. \ref{fg:Image_SNR_vs_Doppler}. It can be seen that in the low Doppler shift region ($\nu < 0.5 $), all three methods output similar image SNRs. As Doppler shift increases, the image SNRs of our time-domain method and the symbol division method decreases severely first and then increases. The minimum SNRs ($<$0 dB) are achieved at integer $\nu$. However, our DAFT-domain method can always output an image SNR greater than 50 dB, which verifies that our DAFT-domain method can work in both integer and fractional normalized Doppler shift scenarios.
 
\begin{figure}
	\centering
	\subfigure[]{
		\includegraphics[width=1.46in]{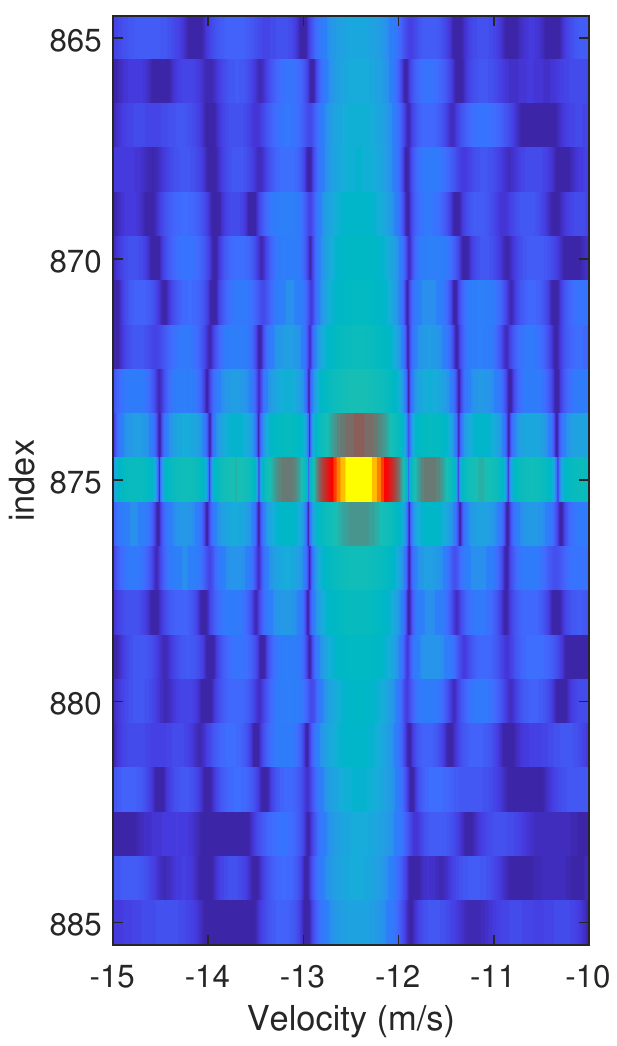}
	}
	\subfigure[]{
		\includegraphics[width=1.7in]{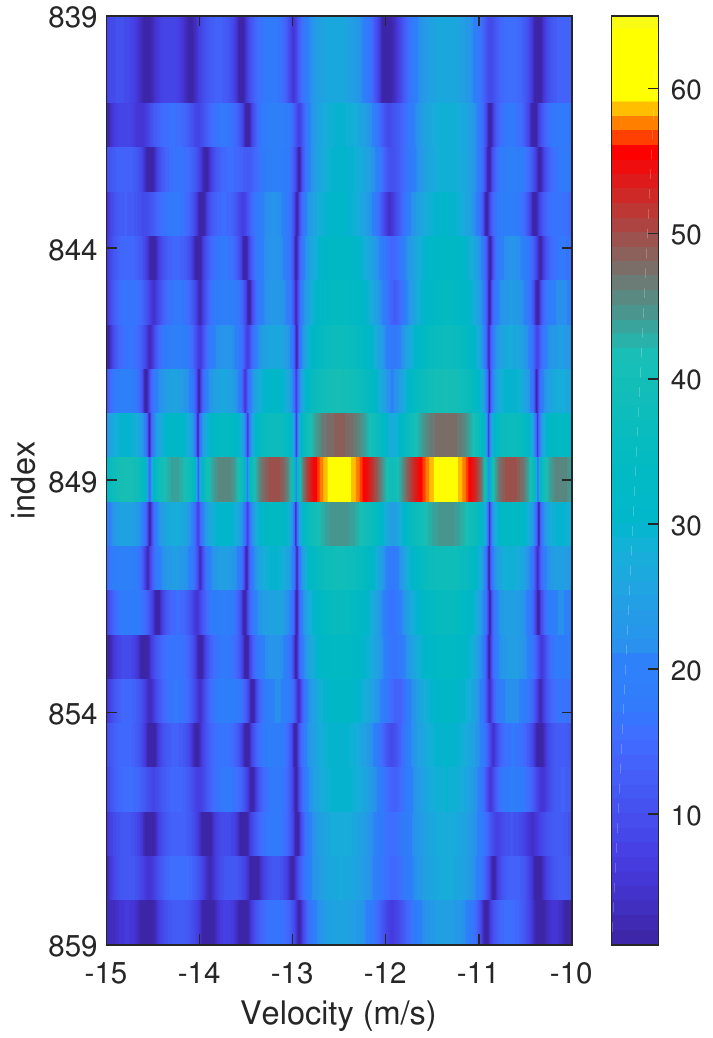}
	}
	\vspace*{-5pt} 
	\caption{Normalized radar images calculated with our AFDM-based DAFT-domain method for 3 targets with ${\rm SNR}_{\rm sig} {=} 0$ dB,  corresponding the range (a) $R{=}$400 m and (b) $R{=}$402 m, respectively.  
		\label{fg:2D_radar_image}}
	\vspace*{-5pt} 
\end{figure}

Figure \ref{fg:2D_radar_image} shows radar image obtained by our DAFT-domain method for multiple targets. There are $P{=}3$ point-like targets with ranges $R {=} \left[400 \ {\rm m}, 402 \ {\rm m}, 402 \ {\rm m}\right]$ and velocities $v_{rel} {=} \left[255 \ {\rm m/s}, 255 \ {\rm m/s}, 256 \ {\rm m/s}\right]$ (corresponding $f_d{=}1.8\Delta_f$). 
Unlike the radar image obtained by the symbol division method \cite{sturm2011waveform}, our DAFT-domain method produces multiple radar image simultaneously. Each radar image contains the information of the targets having the same range. As shown in Figs. \ref{fg:2D_radar_image}(a) and \ref{fg:2D_radar_image}(b), there are two radar images that contain peaks exceeding the threshold. The latter contains two targets with different velocity. Hence, all three targets can be detected from these radar images.

Table \ref{tab:estimate_result} lists the velocity estimation of our time-domain and DAFT-domain methods with $P{=}1$ and $l{=}128$, where $f_d$ and $v_{rel}$ denote truth Doppler shift and corresponding velocity of target, respectively. $\hat v_1$ and $\hat v_2$ are estimated velocities of time-domain and DAFT-domain methods, respectively. We can see that when $f_d {=}0.45\Delta_f$, both methods can correctly estimate the velocity.  When $f_d {=}1.4\Delta_f$ and $f_d {=}2.0\Delta_f$, the time-domain method outputs the wrong velocity due to ambiguity (its maximum unambiguous Doppler shift ${=}\pm{1 \mathord{\left/
{\vphantom {1 {\left( {2{T_{AFDM}}} \right)}}} \right.
\kern-\nulldelimiterspace} {\left( {2{T_{AFDM}}} \right)}} {=} \pm 0.47 \Delta_f$). However, the DAFT-domain method still works well.  

\renewcommand\arraystretch{1.2}
\begin{table}
	\centering
	\caption{Comparison of estimation of velocity}\label{tab:estimate_result}
	\begin{threeparttable}
		\begin{tabular}{|c|c|c|c|}
			\hline
			\textbf{$f_d$}                       &\textbf{Velocity $v_{rel}$}                       & \textbf{Velocity $\hat v_1$} & \textbf{Velocity $\hat v_2$}   \\ \hline
			0.45$\Delta_f$ &  63.9 m/s (230.1 km/h)   & 63.7 m/s & 63.7 m/s    \\ \hline
			1.4$\Delta_f$ &  197.6 m/s (711.5 km/h)   & 63.7 m/s & 197.4 m/s    \\ \hline
			2.0$\Delta_f$ &  284.1 m/s (1022.74 km/h)   & -65.3 m/s & 284.1 m/s    \\ \hline
		\end{tabular}
	\end{threeparttable}
\end{table}

\section{Conclusion}
In this paper, we proposed an AFDM-based ISAC systems. Two methods were proposed to estimate the range and velocity of targets in the time domain and the DAFT domain, respectively. The latter significantly increased the maximum unambiguous Doppler shift and still enjoyed good PSLR and image SNR performances in high mobility scenarios.


\section*{Acknowledgment}
This work was supported by the National Natural Science Foundation
of China under Grant 61971025, 62071026 and 61941106. This research
was supported by the high performance computing (HPC) resources at
Beihang University.



%

%
%


\bibliographystyle{IEEEtran}
\bibliography{IEEEabrv,IEEE_JRCJ_ref}

\end{document}